\documentclass[lettersize,journal]{IEEEtran}
\usepackage{amsmath,amsfonts}
\usepackage{algorithmic}
\usepackage{algorithm}
\usepackage{amssymb}
\usepackage{array}
\usepackage[caption=false,font=footnotesize,labelfont=sf,textfont=sf]{subfig}
\usepackage{booktabs}
\usepackage{caption}
\usepackage{textcomp}
\usepackage{stfloats}
\usepackage{url}
\usepackage{verbatim}

\usepackage{graphicx}
\usepackage[english]{babel} 
\usepackage{lipsum} 

\usepackage{amsthm}

\newtheorem{remark}{Remark}

\usepackage{float}
\usepackage{stfloats}
\usepackage{cite}

\begin{document}



\title{Learnable Kernels for FRI: Joint Kernel–Encoder Optimization and Hardware Validation}
\author{Omkar~Nitsure, Sampath Kumar Dondapati, and Satish Mulleti,~\IEEEmembership{Member,~IEEE}
\thanks{ON, SKD, and SM are with the Department of Electrical Engineering, Indian Institute of Technology (IIT) Bombay, India. SKD is also associated with Satish Dhawan Space Center-SHAR, ISRO, India.
Emails: omkarnitsure2003@gmail.com, mr.sampathkumar.d@gmail.com,
mulleti.satish@gmail.com}}



\maketitle

\begin{abstract}
Finite Rate of Innovation (FRI) sampling techniques provide efficient frameworks for reconstructing signals
with inherent sparsity at rates below Nyquist. However, traditional FRI reconstruction methods rely heavily on pre-defined kernels, often limiting hardware implementation and reconstruction accuracy under noisy conditions. In this paper, we propose a robust, flexible, and practically implementable framework for FRI reconstruction by introducing novel learnable kernel strategies. First, we demonstrate effective reconstruction using known, fixed kernels such as truncated Gaussian and Gaussian pair kernels, which mitigate the requirement that the samples should have a sum-of-exponentials (SoE) form. Next, we extend this concept by jointly optimizing both the sampling kernel and reconstruction encoder through a unified learning approach, yielding adaptive kernels that significantly outperform traditional methods in resolution and noise robustness, with reduced sampling rates. Furthermore, we propose a practical hardware realization by representing kernels as sums of two exponential decay signals with jointly optimized poles, facilitating compact, efficient analog implementations. Our approach is validated experimentally through hardware implementations using a unity-gain Sallen-Key analog filter, achieving accurate real-world signal recovery. The developed convolutional neural network-based encoder substantially reduces computational complexity, demonstrating competitive performance with fewer parameters, making our method particularly suitable for resource-constrained, edge-based deployments.
\end{abstract}

\begin{IEEEkeywords}
Finite rate of innovation, sub-Nyquist sampling, sampling kernel design, learnable filters, FRI hardware demonstration.
\end{IEEEkeywords}

\section{Introduction}
\IEEEPARstart{F}{inite Rate of Innovation} (FRI) sampling offers a sampling-efficient alternative to the Shannon--Nyquist (SN) theorem \cite{nyquist,shannon}, enabling the reconstruction of sparse signals at sub-Nyquist rates. The SN framework applies to bandlimited signals and requires sampling above the Nyquist rate, that is, twice the maximum signal frequency. However, as the sampling rate increases, the power consumption and quantization error of analog-to-digital converters (ADCs) also increase \cite{walden_1999, binle_adc}, making the SN approach unsuitable for wideband applications such as communication and ultra-wideband radar. In contrast, many signals in these applications have only a few degrees of freedom and can be modeled as FRI, allowing the use of low-power ADCs at reduced sampling rates \cite{mulleti2025power}.

\begin{figure}[!t]
    \centering
    \includegraphics[width= 3.5 in]{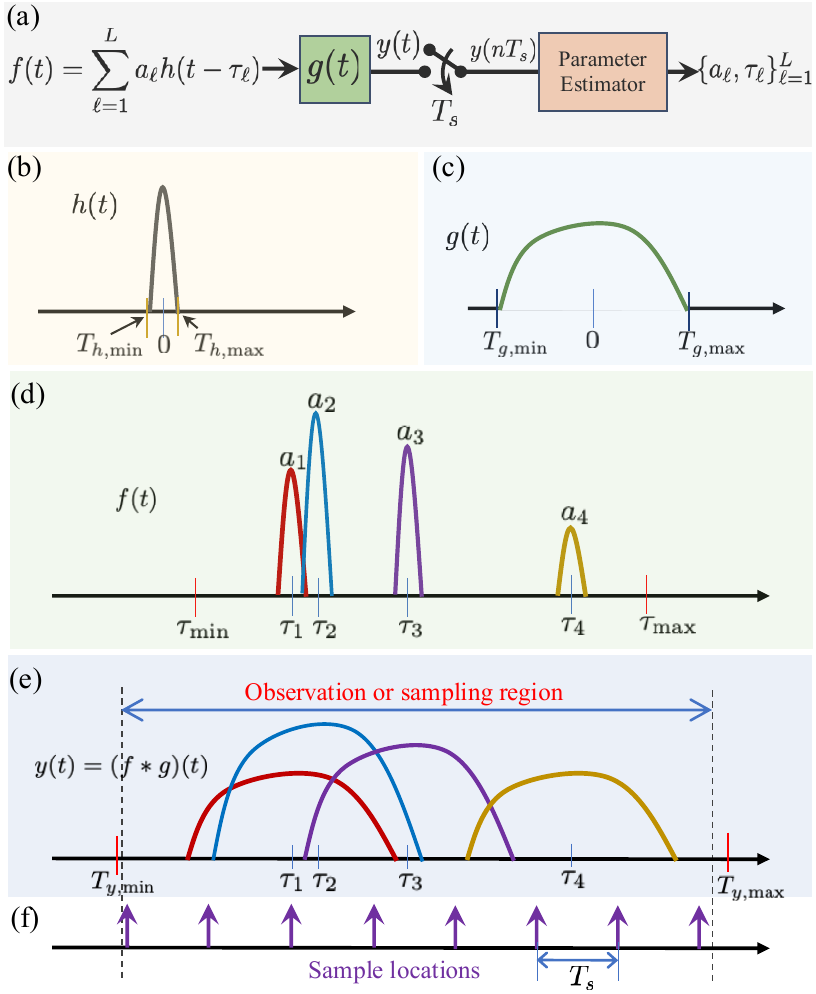}
     \caption{Overview of the proposed FRI pipeline: the signal is prefiltered by a kernel $g_{\boldsymbol{\theta}}$, to increase the spread and reduce the sampling rate.}
    \label{fig:fri_filtering2}
\end{figure}

FRI principles have been successfully applied in diverse domains, including curve fitting \cite{pan_curve, ellipse_tip,fatemi_curve}, radar imaging \cite{bar_radar, bajwa_radar, sunil}, light detection and ranging (LIDAR) \cite{castorena_lidar}, time-domain optical coherence tomography (TDOCT) \cite{blu_oct}, ultrasound imaging \cite{eldar_sos,eldar_beamforming,mulleti_icip}, frequency-domain OCT (FDOCT) \cite{css_fdoct, mulleti_fdoct,mulleti_interleaved}, source localization \cite{bruce_diffuse, dokmanic_tsp}, ECG compression \cite{crochiere, pina_ecg, hao,bachler_vpwfri}, EEG reconstruction \cite{pina_eeg}, intelligent transportation \cite{condat_map}, neural signal processing \cite{onativia}, astronomical signals \cite{pan_redioastronomy}, and MIMO communication systems \cite{barbotin}, among others.

Most applications use a time-of-flight model, where the FRI signal is expressed as a linear combination of time-delayed and amplitude-scaled versions of a known pulse. The key challenge is to estimate the unknown delays and amplitudes from sub-Nyquist samples. Below, we formulate the problem and summarize prior results.

\subsection{Problem Formulation and Existing Results} \label{sec:pf}

Consider a class of FRI signals $C_h$ where a signal $f(t) \in C_h$ is given by
\begin{align}
    f(t) = \sum_{\ell=1}^L a_\ell \, h(t-\tau_\ell), \label{eq:f(t)}
\end{align}
where $h(t)$ is a known pulse, and $a_\ell$, $\tau_\ell$ are the amplitude and delay of the $\ell$-th pulse. We assume:
\begin{itemize}
    \item[\bf A0:] Compact support: $h(t) = 0$ for $t \notin [T_{h,\min},T_{h,\max}]$.
    \item[\bf A1:] Finite amplitude range: $a_\ell \in [a_{\min},a_{\max}]$.
    \item[\bf A2:] Known delay range: $\tau_\ell \in [\tau_{\min},\tau_{\max}]$, with $\tau_1<\tau_2<\cdots<\tau_L$.
    \item[\bf A3:] The model order $L$ is known.
\end{itemize}

The goal of FRI sampling is to estimate $\{a_\ell,\tau_\ell\}_{\ell=1}^L$ from discrete measurements of $f(t)$ such as uniform sampling. However, direct uniform sampling requires $T_s$ smaller than the support of $h(t)$, which can lead to very high rates for wideband pulses. To reduce sampling demands, filtering-based methods are used (Fig.~\ref{fig:fri_filtering2}(a)). A wide-support kernel $g(t)$ spreads each pulse (Fig.~\ref{fig:fri_filtering2}(e)), allowing uniform sampling at lower rates (Fig.~\ref{fig:fri_filtering2}(f)). Since there are $2L$ unknowns, at least $N \geq 2L$ samples are needed. Thus, the sampling kernel plays a central role in FRI reconstruction.

Conventional kernels are chosen so that the samples (or their linear combinations) yield a sum-of-exponentials (SoE) sequence
\begin{align}
  z(n) = \sum_{\ell = 1}^L b_{\ell} \, u_\ell^n, \quad n=0,\dots,N-1, \label{eq:z(n)}
\end{align}
where $u_\ell$ is an invertible function of $\tau_\ell$ (such as $u_\ell=e^{\mathrm{j}\omega_0\tau_\ell}$ with $\omega_0\tau_{\max}<2\pi$), and $b_\ell$ depends on $\{a_\ell,\tau_\ell\}$ \cite{fri_strang,eldar_sos,mulleti_kernal,blu_moms}. High-resolution spectral estimation (HRSE) methods such as annihilating filter \cite{prony}, ESPRIT \cite{paulraj_esprit}, and matrix pencil \cite{sarkar_mp} are then applied. In the noiseless case, $\{a_\ell,\tau_\ell\}$ can be uniquely recovered if $N\geq 2L$.

To generate SoE sequences, specific kernels are used: polynomial/exponential generating functions \cite{fri_strang,blu_moms,bluspmag}, or sum-of-sincs (SoS) kernels \cite{eldar_sos,mulleti_kernal}. Infinite-support kernels such as Gaussian or sinc have also been explored \cite{vetterli}, but they are unstable in noise. A key result is that the minimum sampling rate is $\tfrac{2L}{\tau_{\max}}$ \cite{mulleti_kernal} in the absence of noise, and resolution $\Delta\tau$ is unrestricted, where
\begin{align}
   \Delta \tau = \min_{1 \leq \ell \leq L-1} |\tau_{\ell+1}-\tau_\ell|. \label{eq:resolution}
\end{align}

In noisy settings, annihilating filters become unstable and require preprocessing such as Cadzow denoising \cite{cadzow,condat_cadzow,simeoni_cadzowPnP,cai2015fast}. ESPRIT and matrix pencil improve robustness but degrade as $L$ increases or $\Delta\tau$ shrinks. Sequential estimation \cite{css_sampling,guan2007opportunistic,alexandru2019reconstructing} alleviates complexity but sacrifices resolution.

Learning-based methods have recently emerged \cite{mulleti2023learning,leung2023learning}. In \cite{mulleti2023learning}, SoS kernels were optimized jointly with learned iterative algorithms, but delays were assumed on a grid. The method in \cite{leung2023learning} removed the grid constraint using a deep-unfolding denoiser and an autoencoder (termed as FRIED-Net), but still relied on exponential-generating kernels. Several alternative deep models are also applicable for reconstruction, provided that the samples have an SoE form as in \eqref{eq:z(n)} \cite{izacard2019learning,guo_doa_2020,data_driven_learning,wu_2019, elbir2020deepmusic, xie_damped_2021, pan_learn_2022, chen2022sdoanet, ali_doa_learning_2023,smith2024frequency,biswas2024deep,dondapati2025super}.  

A common limitation remains: these methods depend on fixed kernel families (SoS or exponential-generating). This raises key questions: \emph{Can we learn the sampling kernel itself, jointly with reconstruction? Can such kernels be made hardware-realizable?} These are the challenges we address in this work. Our major contributions are summarized as follows.

\subsection{Our Contributions}
We propose a flexible FRI reconstruction framework that unifies kernel design and estimation. Our main contributions are:

\begin{enumerate}
    \item \textbf{Learnable reconstruction with known kernels:} We show that accurate FRI recovery is possible without SoE-generating kernels. Using truncated Gaussian and Gaussian-pair kernels together with a trained convolutional-neural network (CNN)-based encoder, we achieve stable, noise-robust reconstruction. In particular, a learning-based encoder optimized for a given kernel attains reconstruction accuracy on par with or better than existing methods (see Figs.~\ref{fig:nmse_gaussian_half_ts}--\ref{fig:nmse_gaussian_pair}).
    
    \item \textbf{Joint kernel and reconstruction learning:} We develop a unified framework that learns both the sampling kernel and the encoder via end-to-end training. To obtain a finite parameterization for the continuous-time filter, we represent the kernel as a piecewise-linear (B-spline) function, which provides an expressive yet compact parameterization. Learned kernels exhibit structured shapes (such as truncated-sinc-like or multi-lobe forms) that match signal characteristics. Across a range of signal-to-noise ratios (SNRs) and model orders, the jointly learned system reduces error by about $5$--$6\,$dB relative to FRIED-Net while operating at \(\approx 37\%\) lower sampling rate. (See Figs.~\ref{fig:nmse_learned_smooth}--\ref{fig:nmse_learned_gaussian} and Table~\ref{table:all_kernels}.)
    
     \item \textbf{Hardware design and validation:} The kernels learned in the previous discussions need not be practically realizable. As most analog filters can be designed using resistor-capacitor (RC) circuits together with active components such as operational amplifiers, we are considering optimizing the RC values for a given circuit design. Specifically, we consider a unity-gain Sallen--Key filter which has two poles, which act as learnable parameters. Once the poles, together with the reconstruction network, are optimized for a given set of samples, we implemented the filter in hardware. Our hardware experiments show close agreement with simulations.
    
    \item \textbf{Efficiency and scalability:} We design a lightweight CNN encoder with $0.115$M parameters (60\% fewer than FRIED-Net's $0.287$M) that maintains or improves accuracy across reduced sampling budgets ($N=11$) and higher model orders ($L=5,10$), making the approach suitable for edge deployment (Tables~\ref{table:N_11_samples} and \ref{table:L_5_10_locs}).
\end{enumerate}

Collectively, these contributions yield a practical, noise-robust, and efficient FRI reconstruction pipeline that outperforms prior methods across diverse conditions.

The paper is organized as follows. In the next section, we discuss our main solution for jointly learning the sampling kernel and the reconstruction. In Section \ref{sec:fixed_kernel}, we discuss reconstruction for any kernel. The joint kernel and reconstruction are discussed in Section~\ref{sec:joint_kernel}, whereas Section~\ref{sec:implementable_kernel} discusses a learning method for an implementable kernel. We conclude in Section~\ref{sec:conclusion}.

\section{Proposed Learnable Filters and Reconstruction Method}
We now present our approaches for addressing the limitations of existing FRI reconstruction. Specifically, we propose three variants:  
(i) learning-based reconstruction for a known kernel,  
(ii) joint learning of the sampling kernel and reconstruction method, and  
(iii) joint learning of practically implementable kernels with the reconstruction method.  
Before detailing these approaches, we introduce a unified framework that encompasses them.

Consider the FRI signals in $C_h$ under Assumptions {\bf A0--A3}. Let $g_{\boldsymbol{\theta}}(t)$ be a kernel parameterized by $\boldsymbol{\theta}$. The filtered signal $y(t) = (f*g_{\boldsymbol{\theta}})(t)$ is sampled, and the time delays $\boldsymbol{\tau} = [\tau_1, \tau_2, \cdots, \tau_L]$ are estimated using a deep network encoder $E_{\boldsymbol{\phi}}$ \cite{leung2023learning}. Denote the samples by $\mathbf{y}_{\boldsymbol{\theta}}$, where the subscript highlights dependence on the kernel. The encoder outputs the estimated delays:
\begin{align}
    \hat{\boldsymbol{\tau}} = E_{\boldsymbol{\phi}}(\mathbf{y}_{\boldsymbol{\theta}}).
\end{align}

Once the delays are obtained, the amplitudes $\mathbf{a} = [a_1, a_2, \cdots, a_L]$ are estimated from $\mathbf{y}_{\boldsymbol{\theta}}$ and $\hat{\boldsymbol{\tau}}$. We adopt this two-step procedure because the delays are non-linear functions of the samples, whereas the amplitudes are linearly related. Separating the tasks yields higher reconstruction accuracy with a significantly simpler encoder.

We first consider fine-tuning the encoder for a given kernel (fixed $\boldsymbol{\theta}$). Let the training dataset be
\begin{align}
    \mathcal{D}_{\text{train}} = \{\mathbf{y}_{\boldsymbol{\theta}, i}, \boldsymbol{\tau}_i, \mathbf{a}_i\}_{i=1}^I, \label{eq:data}
\end{align}
where $I$ is the number of examples. For each pair $(\boldsymbol{\tau}_i, \mathbf{a}_i)$, the corresponding FRI signal and filtered samples are generated. The encoder is then trained by solving
\begin{align}
    \min_{\boldsymbol{\phi}} \sum_{i=1}^I\|\boldsymbol{\tau}_i -  E_{\boldsymbol{\phi}}(\mathbf{y}_{\boldsymbol{\theta},i})  \|_p^p, \label{eq:opt_encoder} \tag{$\text{P}_{\text{Encoder}}$}
\end{align}
with $p=1$ in all our implementations.

Next, we extend this framework to jointly learn the kernel and the encoder by solving
\begin{align}
    \min_{\boldsymbol{\theta, \phi}} \sum_{i=1}^I\|\boldsymbol{\tau}_i -  E_{\boldsymbol{\phi}}(\mathbf{y}_{\boldsymbol{\theta},i}) \|_p^p, \label{eq:opt_joint} \tag{$\text{P}_{\text{Joint}}$}
\end{align}
where the kernel is parameterized either as an arbitrarily shaped function or in a hardware-realizable form. Both (\ref{eq:opt_encoder}) and (\ref{eq:opt_joint}) are solved using backpropagation with automatic differentiation. At iteration $k$, the updates are
\begin{equation}
\boldsymbol{\phi}^{(k+1)} = \boldsymbol{\phi}^{(k)} - \eta_{\phi}
\displaystyle\sum_{i} \nabla_{\boldsymbol{\phi}} \left\| \boldsymbol{\tau}_i -
E_{\boldsymbol{\phi}}(\mathbf{y}_{\boldsymbol{\theta}, i}) \right\|_p^p
\end{equation}
\begin{equation}
\boldsymbol{\theta}^{(k+1)} = 
\begin{cases}
\boldsymbol{\theta}^{(k)}, \qquad  \text{for (\ref{eq:opt_encoder})},& \\[4pt]
\boldsymbol{\theta}^{(k)} - \eta_{\boldsymbol{\theta}}
\displaystyle\sum_{i} \nabla_{\boldsymbol{\theta}} \left\| \boldsymbol{\tau}_i -
E_{\boldsymbol{\phi}}(\mathbf{y}_{\boldsymbol{\theta}, i}) \right\|_p^p, & \text{for (\ref{eq:opt_joint})}
\end{cases}
\end{equation}
where $\eta_{\boldsymbol{\phi}}$ and $\eta_{\boldsymbol{\theta}}$ are the respective learning rates.

After obtaining $\hat{\boldsymbol{\tau}}$, amplitudes are estimated by minimizing the mean-squared error (MSE) between the observed and model-predicted samples: $\hat{\mathbf{a}} = \arg\min_{\mathbf{a}} \mathcal{L}_{\text{MSE}}$, where $\displaystyle \mathcal{L}_{\text{MSE}} = \frac{1}{N} \sum_{n=1}^{N} \left| \hat{y}(nT_s; \mathbf{a}, \hat{\boldsymbol{\tau}}, g_{\boldsymbol{\theta}}) - \mathbf{y}_{\boldsymbol{\theta}}[n] \right|^2$ and
$\hat{y}(nT_s; \mathbf{a}, \hat{\boldsymbol{\tau}}, g_{\boldsymbol{\theta}})$ denotes the predicted samples. In practice, $\mathbf{a}$ is initialized with Gaussian random values and refined iteratively using gradient descent:
\begin{align}
    \mathbf{a}^{(k+1)} = \mathbf{a}^{(k)} - \eta \nabla_{\mathbf{a}} \mathcal{L}_{\text{MSE}}(\mathbf{a}^{(k)}).
\end{align}
This procedure exploits the differentiability of the learned signal model, yielding accurate amplitude recovery even in noisy settings.

\begin{remark} \label{remark:Ts}
Unlike conventional FRI methods with fixed sampling rates, our framework allows user-specified $T_s$. For a given $T_s$, the kernel support is chosen such that $T_{g,\max} - T_{g,\min} > T_s$, ensuring sufficient information from each filtered pulse (cf. Section~\ref{sec:pf}).
\end{remark}

In the next section, we first analyze the proposed approach for a fixed kernel.

\section{Learnable Reconstruction For A Known Kernel}\label{sec:fixed_kernel}
We consider a learning-based reconstruction method where the samples are generated from an arbitrary kernel. Our objective is to show that the sampling kernel need not belong to the class that yields SoE sequences. To this end, we use truncated Gaussian and Gaussian-pair kernels. Note that, unlike \cite{vetterli}, which uses Gaussian kernels together with annihilating-filter reconstruction to produce SoE sequences, we do not attempt to form SoEs. Consequently, our approach avoids the instability that Gaussian-based SoE formation exhibits in noisy settings.

To assess effectiveness, we compare against the FRIED-Net framework \cite{leung2023learning}, which has two variants: an encoder-only architecture (referred to as \emph{encoder-of-FRIED-Net}) and an encoder-decoder variant (\emph{FRIED-Net}). For fair comparison, we adopt the same simulation settings (number of samples, amplitude, and delay ranges) as in \cite{leung2023learning}. While \cite{leung2023learning} focuses on periodic FRI signals, we consider aperiodic signals that better reflect many real applications. We retrain FRIED-Net on our aperiodic dataset so both models are evaluated under identical conditions; thus, observed differences arise from model design and learning strategy.

\subsection{Gaussian Kernels}
\label{sec:Gaussain Kernels}
We consider two compactly supported kernels over $[T_{g,\min}, T_{g,\max}]$: a truncated Gaussian
\begin{align}
g(t) = \exp\!\left(-\tfrac{t^{2}}{2\sigma^{2}}\right), \label{eq:trunc_gauss}
\end{align}
and a pair of truncated Gaussians
\begin{align}
g(t) = A \exp\!\left(-\tfrac{(t-t_{1})^{2}}{2\sigma^{2}}\right)
     + B \exp\!\left(-\tfrac{(t-t_{2})^{2}}{2\sigma^{2}}\right).
\end{align}
For both the kernels, we have that $g(t) = 0$ for $t \not \in [T_{g,\min}, T_{g,\max}]$.
For the kernels, $\sigma$ controls the spread, $A,B\in\mathbb{R}$, and centers $t_1,t_2$ are chosen so both lobes lie well within the truncation window (Parameter choices appear in the experiments.). While other shapes are possible, Gaussians are convenient due to closed-form expressions used during sample generation.

\begin{figure}[!t]
    \centering
    \includegraphics[width=3.4in]{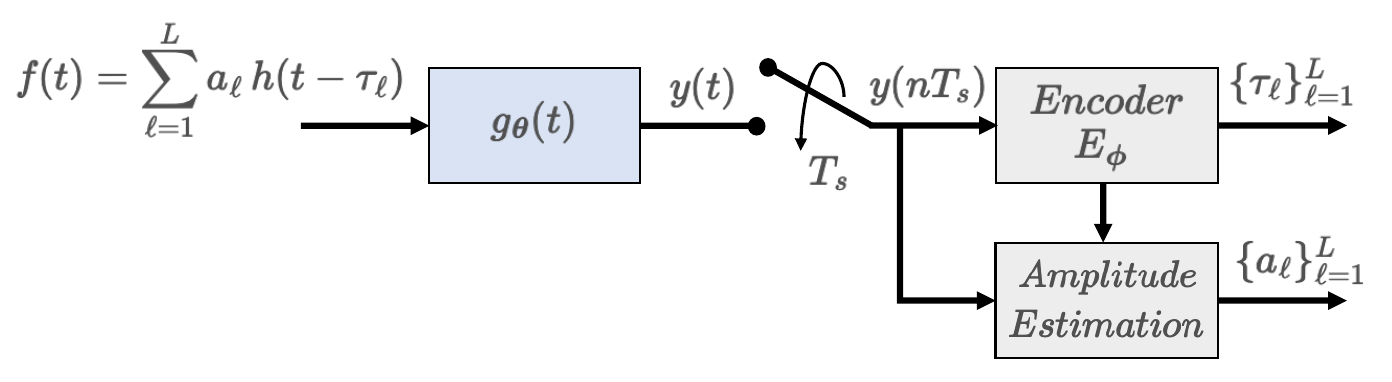}
    \caption{Block diagram of the joint kernel + reconstruction pipeline: input FRI signal $f(t)$ is filtered, uniformly sampled, and decoded by the encoder to estimate $\{\tau_\ell,a_\ell\}$.}
    \label{fig:block_diagram}
\end{figure}

\subsection{FRI Data Generation}\label{sec:Data_gen}

For training and evaluation, we synthesize FRI signals
\[
f(t) = \sum_{\ell=1}^{L} a_\ell \,\delta(t-\tau_\ell),
\]
with amplitudes $a_\ell$ and delays $\tau_\ell$ drawn uniformly from $[a_{\min}, a_{\max}]$ and $[\tau_{\min}, \tau_{\max}]$, respectively. We compare our methods with FRIED-Net for $L=2$, studying resolution $\Delta\tau$ under varying noise levels. Additive noise is applied to the samples to obtain a targeted signal-to-noise ratio (SNR).

In FRIED-Net, a compactly supported exponential-generating kernel is used with support equal to the delay interval: $[T_{g,\min}, T_{g,\max}] = [\tau_{\min},\tau_{\max}]$. Then $N$ samples of the filtered signal $y(t)$ are measured over $[\tau_{\min},\tau_{\max}]$. The observation interval is thus smaller than the full support of $y(t)$, which is
$[T_{y,\min},T_{y,\max}] = [T_{g,\min}+\tau_{\min},\,T_{g,\max}+\tau_{\max}]$ \cite{leung_fri_learn_2023}.

By contrast, our truncated Gaussian filters need not match $[\tau_{\min},\tau_{\max}]$; we take the same number of $N$ samples across the entire support of the filtered signal. Hence, the effective sampling interval in our setup is larger than that in \cite{leung_fri_learn_2023}.

For both FRIED-Net and our Gaussian-based methods, we generated one million training examples spanning diverse amplitudes and delays. For each example, we produced filtered samples and added noise according to the desired SNR. For testing, we generated 1000 examples for each $(\Delta\tau,\text{SNR})$ pair to evaluate resolution performance, and an additional 1000 random examples to assess general performance.

\begin{figure}[!t]
\centering
\includegraphics[width=2in]{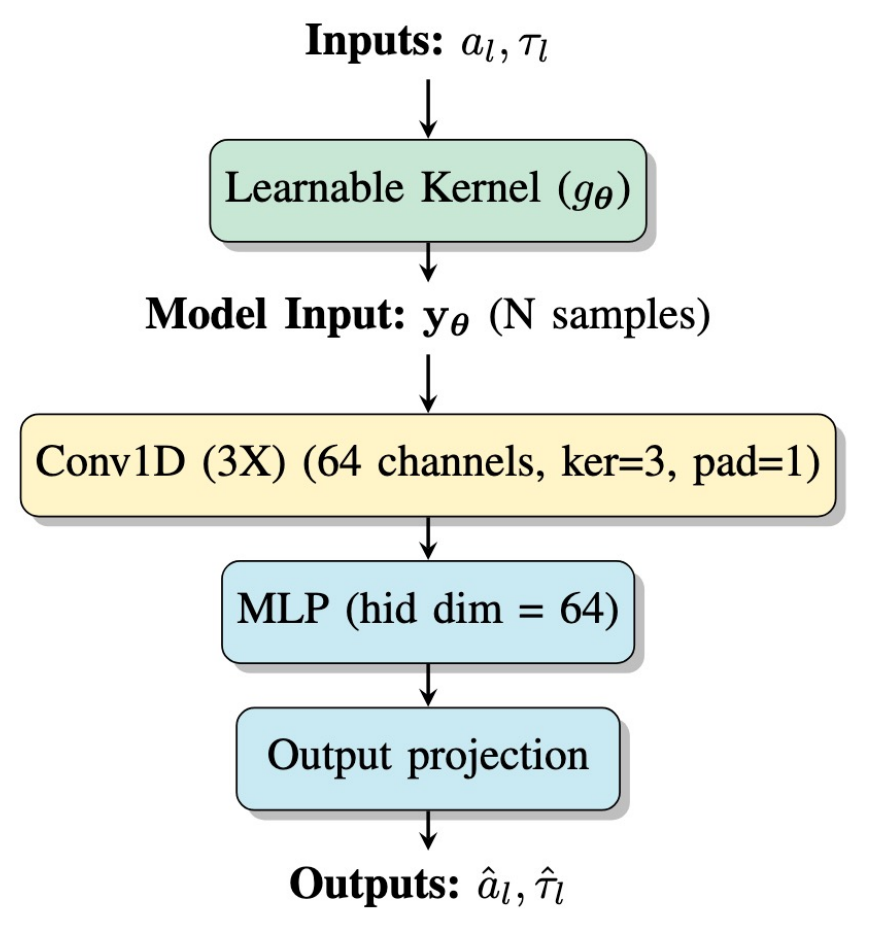}
\caption{Encoder $E_{\boldsymbol{\phi}}$: a compact 1-D Conv + MLP stack mapping sampled input $\mathbf{y}_{\boldsymbol{\theta}}$ to delay estimates $\hat{\boldsymbol{\tau}}$ (amplitudes via least squares).}
\label{fig:arch}
\end{figure}

\subsection{Encoder Architecture and Training}
The encoder $E_{\boldsymbol{\phi}}$ is a CNN that maps $\mathbf{y}_{\boldsymbol{\theta}}\in\mathbb{R}^N$ to delay estimates $\boldsymbol{\tau}\in\mathbb{R}^L$. Each convolution is followed by a GELU activation to introduce smooth nonlinearity; the final layer has no activation to allow real-valued outputs. Fig.~\ref{fig:arch} shows the architecture. Training uses an $\ell_1$ loss, and parameters $\boldsymbol{\phi}$ are optimized with AdamW and cosine-annealing learning-rate scheduling. In joint training, gradients also update kernel parameters $\boldsymbol{\theta}$.

Training details: 2000 epochs, batch size 8192; data is shuffled each step to improve generalization.

\begin{figure}[!t]
\centering
\includegraphics[width=2.2in]{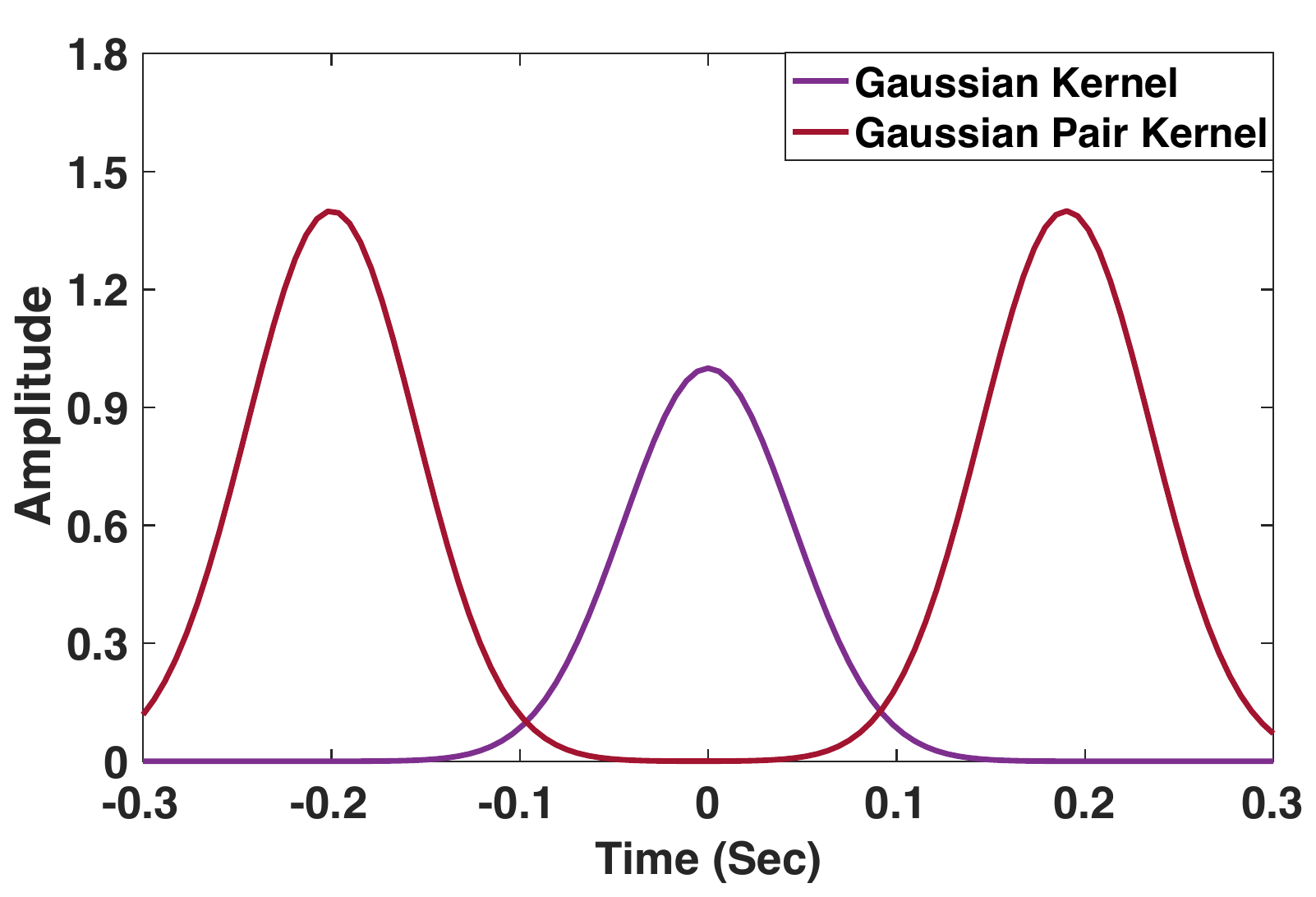}
\caption{Gaussian kernels used: single Gaussian ($\sigma=T_s/2$) and Gaussian pair (dual peaks).}
\label{fig:gaussian_kernels}
\end{figure}

\subsection{Results and Discussion}\label{sec:fixed_results}
We compare FRIED-Net (encoder-only and encoder-decoder) and our Gaussian-based encoders for $L=2$ with
\begin{align}
\begin{aligned}
        [a_{\min}, a_{\max}] &= [0.5,10], \\
        [\tau_{\min}, \tau_{\max}] &= [-0.48,0.52].
\end{aligned}
\label{eq:a_tau_range}
\end{align}
We used $N=21$ samples as in \cite{leung_fri_learn_2023}, which corresponds to $T_s = 0.048$ for the FRIED-Net methods. For the Gaussian filters (Section~\ref{sec:Gaussain Kernels}) we set $\sigma=0.038$, $A=B=1.4$, $t_1=-0.2$, $t_2=0.2$, and support
\begin{align}
[T_{g,\min},T_{g,\max}] = [-0.3,0.3]. \label{eq:gaussian_support}   
\end{align}
These filters are depicted in Fig.~\ref{fig:gaussian_kernels}. Since we sample across the full support of $y(t)$, the effective sampling interval in our experiments is $T_s=0.076$, that is, $\approx 37\%$ reduction in the sampling rate used by FRIED-Net under matched conditions. The choices also ensure $T_s < T_{g,\max}-T_{g,\min}$ (Remark~\ref{remark:Ts}).

\begin{figure}[!t]
\centering
\includegraphics[width=3.4in]{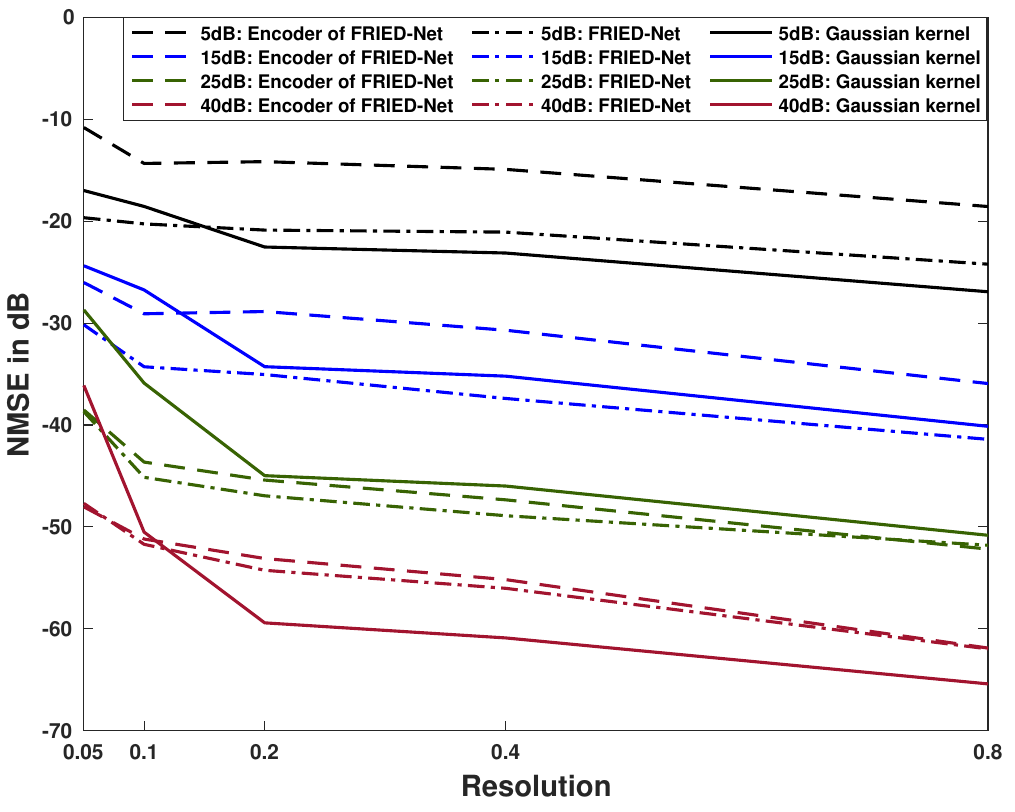}
\caption{NMSE comparison: Gaussian kernel $(\sigma=T_s/2)$ vs FRIED-Net (encoder-only and encoder-decoder) across SNRs 5–40 dB.}
\label{fig:nmse_gaussian_half_ts}
\end{figure}
\begin{figure}[!t]
\centering
\includegraphics[width=3.4in]{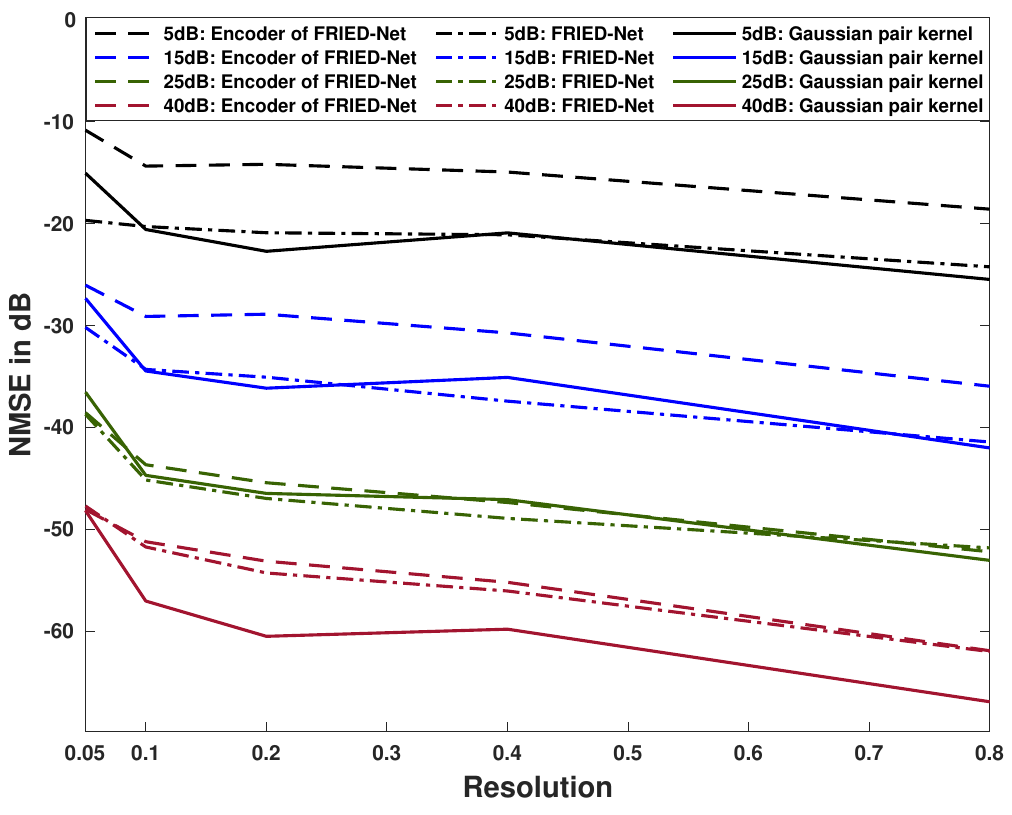}
\caption{NMSE comparison: Gaussian-pair kernel outperforms encoder-only baseline, particularly at low SNR.}
\label{fig:nmse_gaussian_pair}
\end{figure}

Figs.~\ref{fig:nmse_gaussian_half_ts} and \ref{fig:nmse_gaussian_pair} summarize performance across SNRs in terms of normalized MSE (NMSE) measured as
\begin{align}
    \text{NMSE (in dB)} = 10\, \log_{10} \left(\frac{ \|\mathbf{\tau} -  \hat{\mathbf{\tau}}\|^2}{\|\mathbf{\tau}\|^2}\right), \notag 
\end{align}
where $\hat{\mathbf{\tau}}$ is an estimate of the delays $\mathbf{\tau}$.
At 5\,dB the single Gaussian achieves a $2$–$4$\,dB NMSE improvement over the FRIED-Net encoder; the Gaussian pair yields $5$–$7$\,dB gains over the encoder and $1$–$2$\,dB over FRIED-Net. At 15\,dB, the Gaussian pair still outperforms the encoder, though it may lie slightly below FRIED-Net. At high SNRs (25–40\,dB), FRIED-Net narrows the gap and can surpass the fixed Gaussian designs; nonetheless, both Gaussian variants consistently beat the encoder-only baseline.

These results indicate that accurate FRI parameter estimation does not require SoE-generating kernels: compact Gaussian-based kernels produce stable, noise-robust recovery. However, fixed, manually chosen kernels may not generalize across signal classes or noise regimes. This observation motivates the next stage: jointly learning both the sampling kernel and the reconstruction network.

\section{Jointly Learned Kernel and Reconstruction}\label{sec:joint_kernel}
In this section, we propose a unified framework that jointly learns the sampling kernel and the encoder. By allowing the kernel shape to adapt to training data, the model can discover sampling kernels that are well matched to the underlying signal structure and the encoder architecture.

\subsection{Kernel Parameterization}
In the proposed joint learning framework, the sampling kernel $ g_{\boldsymbol{\theta}}(t)$ is parameterized as a weighted sum of first-order B-spline basis functions\footnote{$\beta_1(t) = 1 - |t|$, for $|t| \leq 1$, and zero otherwise.}:
\begin{align}
    g_{\boldsymbol{\theta}}(t) = \sum_{k = -K}^{K} c_k \, \beta_1\left( \frac{t - kT}{T} \right), \label{eq:spline_kernel}
\end{align}
where $\boldsymbol{\theta} = \{c_k\}_{k = -K}^{K}$ are the learnable kernel coefficients. This construction yields a piecewise-linear kernel with compact support on $[-KT,KT]$. The spacing parameter $T$ controls resolution and overlap between adjacent basis functions, ensuring that the kernel remains smooth and expressive while keeping the overall support localized. The B-spline basis is compact and differentiable, which enables efficient backpropagation with respect to each coefficient $c_k$.

\begin{remark}
   Alternative basis functions may be used in \eqref{eq:spline_kernel} provided the kernel retains a finite number of learned parameters and allows gradient-based training.
\end{remark}

\begin{figure}[!t]
\centering
\includegraphics[width=2.5in]{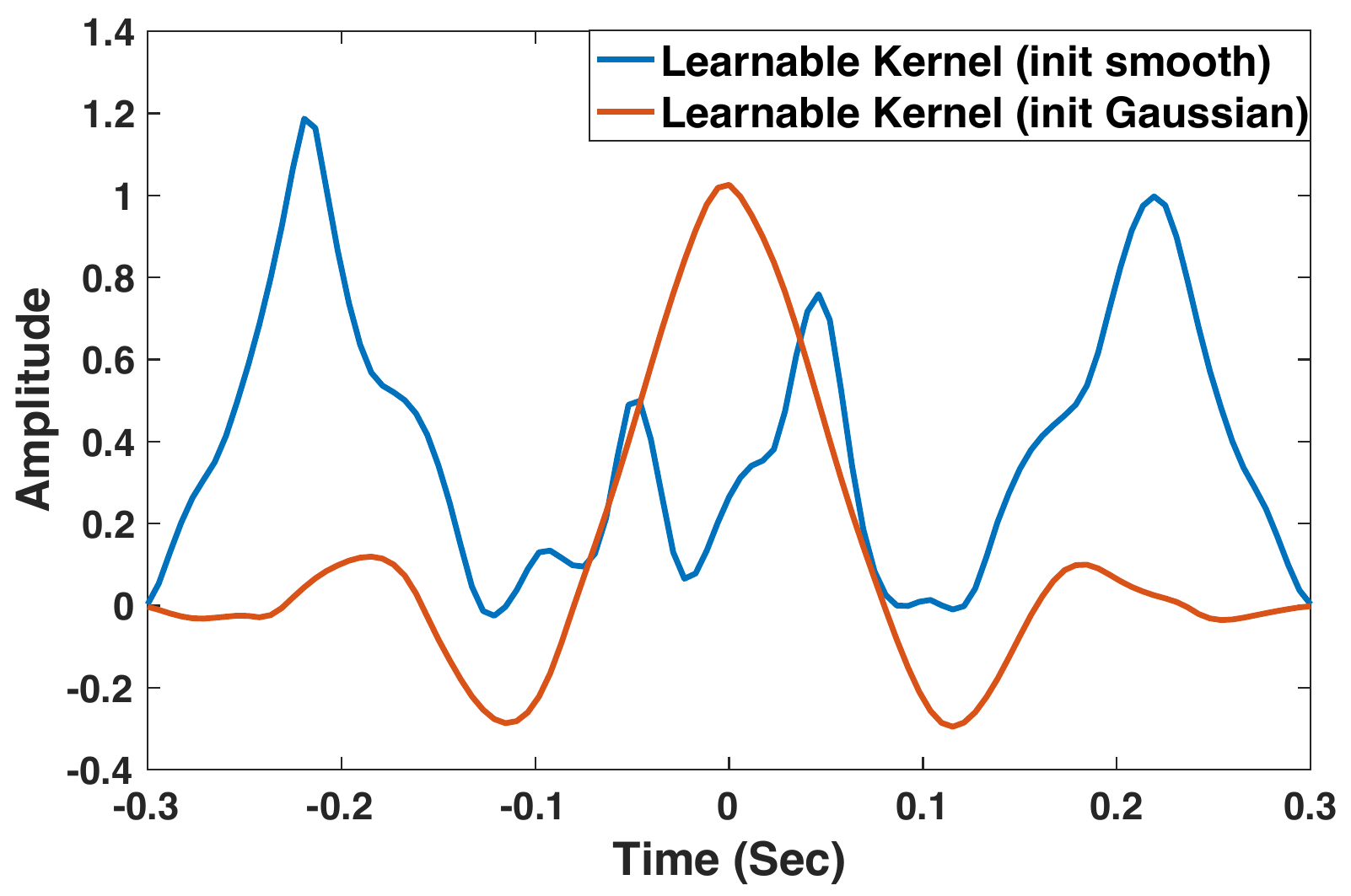}
\caption{Adaptively learned kernel shapes obtained via joint optimization of the kernel coefficients and the encoder $E_{\boldsymbol{\phi}}$, comparing initializations with smooth (low-variance) coefficients and Gaussian-shaped coefficients $(\sigma = T_s/2)$.}
\label{fig:learned_kernels}
\end{figure}

\begin{figure}[!t]
\centering
\includegraphics[width=3.4in]{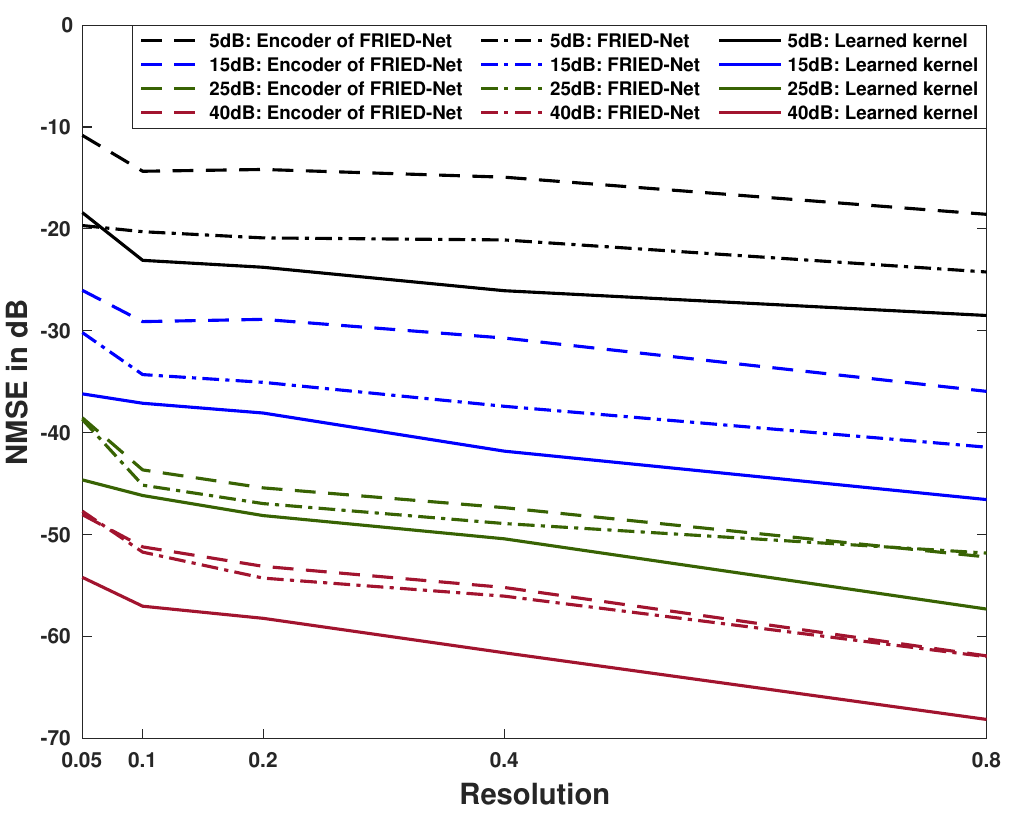}
\caption{NMSE performance: jointly learned kernel (init smooth) yields substantial reconstruction improvement versus FRIED-Net and its encoder across SNRs 5–40 dB. (NMSE reported in dB; see text for definition.)}
\label{fig:nmse_learned_smooth}
\end{figure}

\begin{figure}[!t]
\centering
\includegraphics[width=3.4 in]{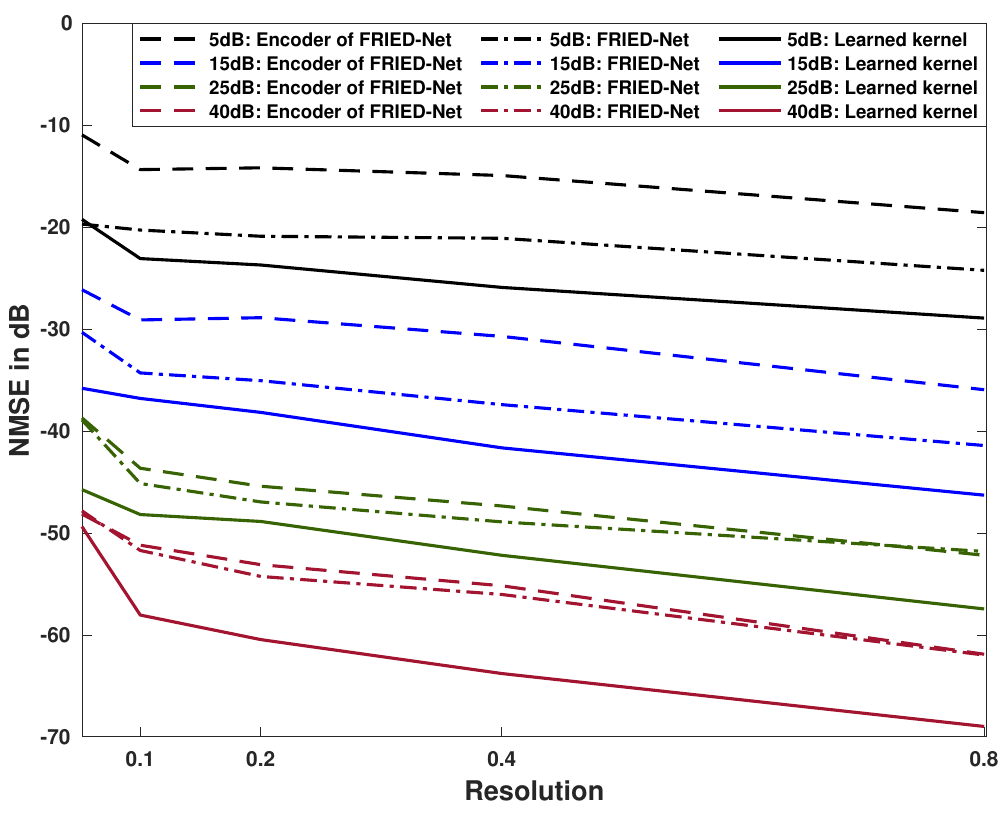}
\caption{NMSE performance: jointly learned kernel (init Gaussian) shows robust improvements over FRIED-Net across SNRs 15–40 dB.}
\label{fig:nmse_learned_gaussian}
\end{figure}

Once the kernel is parameterized, samples are generated as described in Section~\ref{sec:Data_gen}. Joint training of filter parameters and the encoder minimizes the cost in \eqref{eq:opt_joint} with $p=1$. Experiments reported here use $L=2$, $N=21$, $T_s = 0.076$, $K = 52$, and the data ranges in \eqref{eq:a_tau_range}; the kernel support uses \eqref{eq:gaussian_support}. Fig.~\ref{fig:learned_kernels} shows the evolution of learned kernels under different initialization strategies. When initialized as a truncated Gaussian with $\sigma=T_s/2$, the learned kernel tends toward a truncated-sinc-like shape; when initialized smoothly, the learned kernel can become multi-lobed, resembling the two-Gaussian model discussed earlier. These learned shapes are structured and interpretable rather than arbitrary.

\begin{table*}[!t]
  \centering
  \caption{Normalized MSE of various kernels for 1000 test examples ($N=21$ samples). NMSE is reported in dB (see main text).}
  \label{table:all_kernels}
  \renewcommand{\arraystretch}{1.2}
  \setlength{\tabcolsep}{3pt}
  \small
  \begin{tabular}{c *{6}{c}}
    \toprule
    SNR (dB) & Encoder of FRIED-Net & FRIED-Net & Gaussian kernel &
    Gaussian pair & Learnable (init smooth) & Learnable (init Gaussian) \\
    \midrule
    5  & -14.8 & -21.6 & -22.6 & -20.1 & -24.9 & -25.2 \\
    15 & -29.6 & -36.7 & -33.6 & -32.7 & -39.7 & -40.2 \\
    25 & -45.0 & -46.7 & -43.3 & -43.1 & -49.6 & -50.6 \\
    40 & -53.2 & -54.3 & -56.7 & -56.0 & -59.5 & -60.5 \\
    \bottomrule
  \end{tabular}
\end{table*}

\begin{table}[!t]
    \centering
    \caption{Normalized MSE of learnable kernel (init Gaussian) for $N = 11$ and $21$ (time $t_k$ and amplitude $a_k$). NMSE shown in dB.}
    \label{table:N_11_samples}
    \renewcommand{\arraystretch}{1.2}
    \setlength{\tabcolsep}{6pt}
    \begin{tabular}{ccccc}
        \toprule
        SNR (dB) & \multicolumn{2}{c}{$N = 21$} & \multicolumn{2}{c}{$N = 11$} \\
        & $t_k$ & $a_k$ & $t_k$ & $a_k$ \\
        \midrule
        5 & -25.2 & -14.6 & -17.7 & -11.7 \\
        15 & -40.2	& -23.8 & -31.1 & -20.7 \\
        25 & -50.6	& -33.3 & -42.5 & -29.1 \\
        40 & -60.5 & -45.1 & -50.9 & -39.2 \\
        \bottomrule
    \end{tabular}
\end{table}

\subsection{Results and Discussion on Joint Learning}
For all experiments below, FRI data are generated as before using the ranges in \eqref{eq:a_tau_range}. NMSE values are reported in dB.

\subsubsection{Resolution Analysis}
We first evaluate resolution with $L=2$ and $N=21$ samples, using controlled pulse separations $\Delta\tau\in[0.05,0.1]$. Both learned kernels (smooth initialization and Gaussian initialization) were trained under the same conditions and operate at a 37\% lower sampling rate than FRIED-Net (see Sec.~\ref{sec:fixed_results}). The NMSE results are shown in Figs.~\ref{fig:nmse_learned_smooth} and \ref{fig:nmse_learned_gaussian}. Quantitatively, the smooth-initialized kernel improves NMSE by about 6–8\,dB relative to the FRIED-Net encoder and 2–3\,dB relative to FRIED-Net across SNRs. The Gaussian-initialized kernel achieves larger gains in this setup, with up to 9\,dB improvement over the encoder and 4–6\,dB over FRIED-Net. 

\subsubsection{Generalization Accuracy}
We assess generalization on 1000 randomly generated FRI signals with $L=2$ and $N=21$, where amplitudes and delays follow \eqref{eq:a_tau_range} without constraints on $\Delta\tau$. Table~\ref{table:all_kernels} summarizes NMSE (dB) for time-delay estimates $t_k$. Across SNRs, the jointly learned kernels result in lower NMSE. In particular, the Gaussian-initialized learned kernel yields the best overall performance for $t_k$. Quantitatively, the learned kernels have $4-5$ dB lower NMSE compared to the FRIED-Net approach, despite operating at a lower sampling rate.

\subsubsection{Reduced Sampling Rate}\label{par:11_samples}
We retrained the joint kernel–encoder model for $N=11$ samples (halving the number of samples) while keeping the kernel support and observation interval unchanged. Table~\ref{table:N_11_samples} reports NMSE for time delays and amplitudes. Relative to the $N=21$ case, NMSE degrades by approximately 7–8\,dB, yet the $N=11$ model continues to produce accurate parameter estimates across SNRs, demonstrating robustness to reduced sampling budgets.

\subsubsection{High Model Order}
We evaluate scalability at higher model orders using the same kernel supports and data ranges. Two settings are considered: (i) $L=5$, $N=42$ (our $T_s=0.038$ vs FRIED-Net $T_s=0.023$) and (ii) $L=10$, $N=84$ (our $T_s=0.019$ vs FRIED-Net $T_s=0.012$). Table~\ref{table:L_5_10_locs} reports NMSE for these settings. At moderate SNR (10\,dB), the learned kernel yields $6$ dB gain over FRIED-Net for $L=5$. At this SNR, the FRIED-Net approach fails to recover the time delays when $L=10$, whereas the proposed method is able to estimate with an NMSE of $-21.3$ dB. At high SNR (40\,dB), the gain of the proposed method is $5-6$ dB. 

\begin{table}[t]
    \centering
    \caption{Normalized MSE of learnable kernel (init Gaussian) for $L = 5$ and $L = 10$ (mean NMSE in dB).}
    \label{table:L_5_10_locs}
    \renewcommand{\arraystretch}{1.2}
    \setlength{\tabcolsep}{6pt}
    \begin{tabular}{ccccc}
        \toprule
        SNR (dB) & \multicolumn{2}{c}{Learnable-kernel} & \multicolumn{2}{c}{FRIED-Net} \\
        & $L=5$ & $L=10$ & $L=5$ & $L=10$ \\
        \midrule
        10  & -21.3 & -16.8 & -15.3  & -0.02 \\
        40  & -46.5  & -35.1 & -39.0 & -30.1\\
        \bottomrule
    \end{tabular}
\end{table}

\subsection{Computational Complexity}
The encoder backbone is a three-layer CNN with MLP and projection layers. The model has 0.115M parameters, substantially smaller than FRIED-Net's 0.287M parameters. This reduces VRAM footprint and speeds inference, making the approach attractive for edge deployment. All experiments reported here were run on an NVIDIA GeForce RTX 5090 (32\,GB VRAM).

In summary, jointly learning the sampling kernel and the reconstruction network yields resolution-aware, noise-robust recovery with lower sampling rates and reduced model complexity relative to FRIED-Net. The next section discusses parameterizing learned kernels in forms suitable for analog implementation and validates one such mapping experimentally.

\section{Jointly Learned Implementable Kernels With Reconstruction}\label{sec:implementable_kernel}
\begin{figure}[!t]
    \centering
    \includegraphics[width=2.5in]{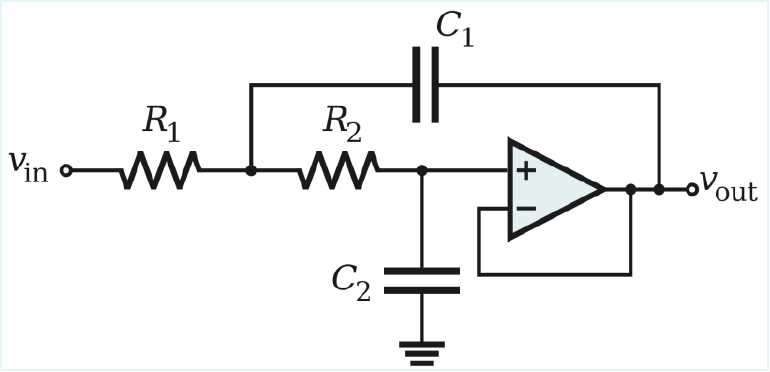}
    \caption{Unity-gain Sallen--Key circuit realizing the learned two-exponential kernel for practical analog implementation.}
    \label{fig:HW_ckt}
\end{figure}

In this section, we elaborate on translating jointly learned kernels to practically realizable analog filters by optimizing circuit hyperparameters. We focus on a second-order unity-gain Sallen--Key analog low-pass filter (Fig.~\ref{fig:HW_ckt}), which realizes a two-pole kernel with impulse response
\begin{equation}
h_{\boldsymbol{\theta}}(t) = A_{0}\left(e^{-\alpha_1 t} - e^{-\alpha_2 t}\right)u(t),
\end{equation}
where $\alpha_1$ and $\alpha_2$ are exponential decay coefficients and the gain factor is set as $A_{0} = \frac{\alpha_1\alpha_2}{(\alpha_2 - \alpha_1)}$ to obtain unity gain. Thus $\boldsymbol{\theta}=[\alpha_1,\alpha_2]$ are the learnable parameters optimized via \((\text{P}_{\text{Joint}})\). Note that resistor-capacitor-based kernels were also suggested in \cite{css_sampling}, with a focus on sequential reconstruction rather than hardware realization. 

The learned poles map to Sallen--Key component values through the standard relations:
\begin{align}
    \begin{aligned}
       \frac{1}{\alpha_1} + \frac{1}{\alpha_2} &=  C_2 (R_1 + R_2), \\
      \frac{1}{\alpha_1 \alpha_2} &= C_1C_2R_1R_2.
    \end{aligned} \label{eq:ckt_param}
\end{align}
Given convenient capacitor choices ($C_1,C_2$), Eqs.~\eqref{eq:ckt_param} yield a quadratic system for $R_1,R_2$ that can be solved in closed form; this gives a direct mapping from the learned continuous-time poles to realizable resistor values. During learning, we constrain the poles, as $\alpha_{\min}\le\alpha_1<\alpha_2\le\alpha_{\max}$, to keep mapped component values within practical ranges.

\begin{figure}[!t]
    \centering
    \includegraphics[width=3.2in]{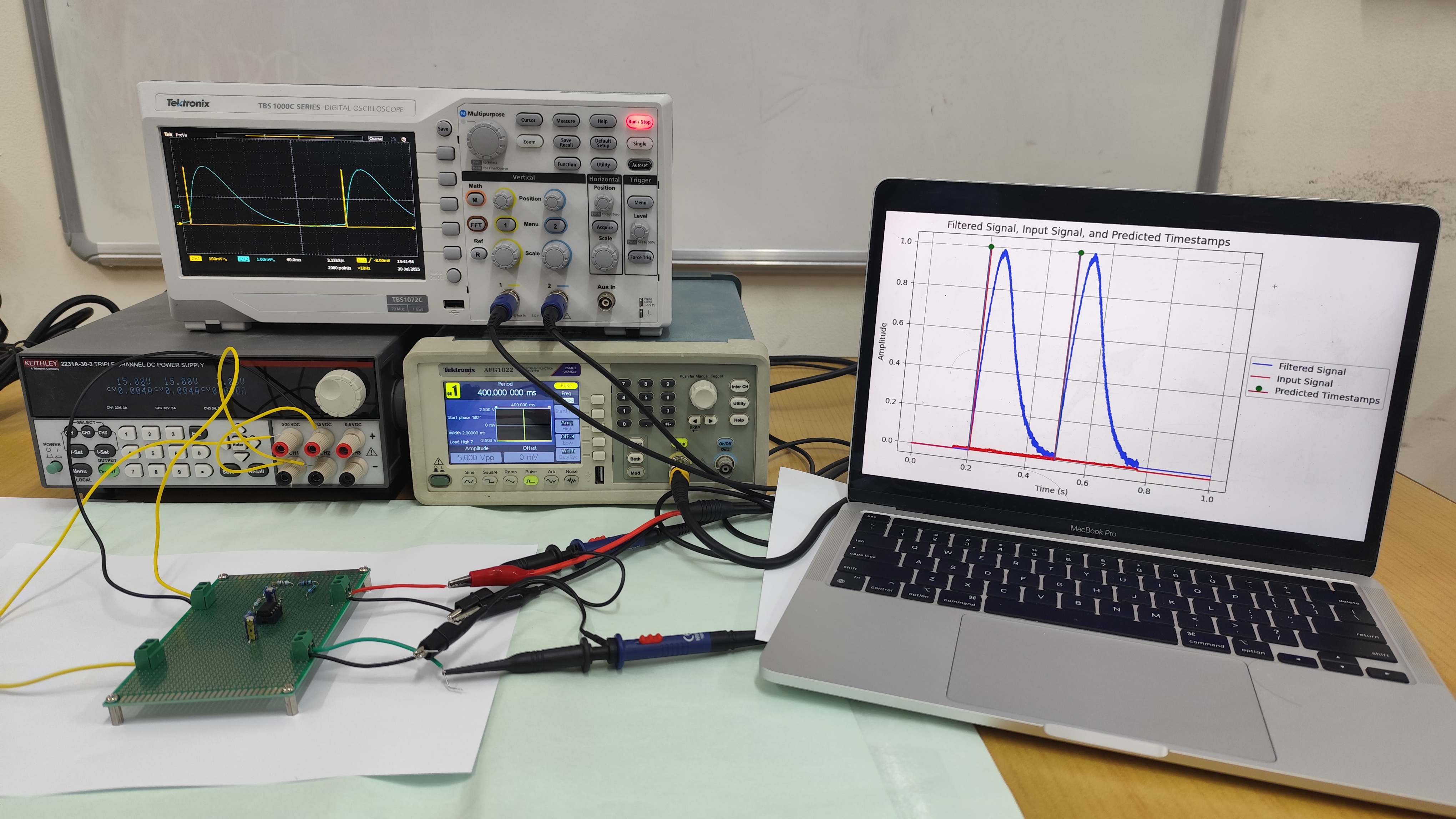}
    \caption{Experimental bench setup: AWG, the Sallen--Key filter board, dual-rail power supply, and DSO connections used for capturing the filtered waveform.}
    \label{fig:HW_stp}
\end{figure}

\begin{figure}[!t]
    \centering
    \includegraphics[width=3in]{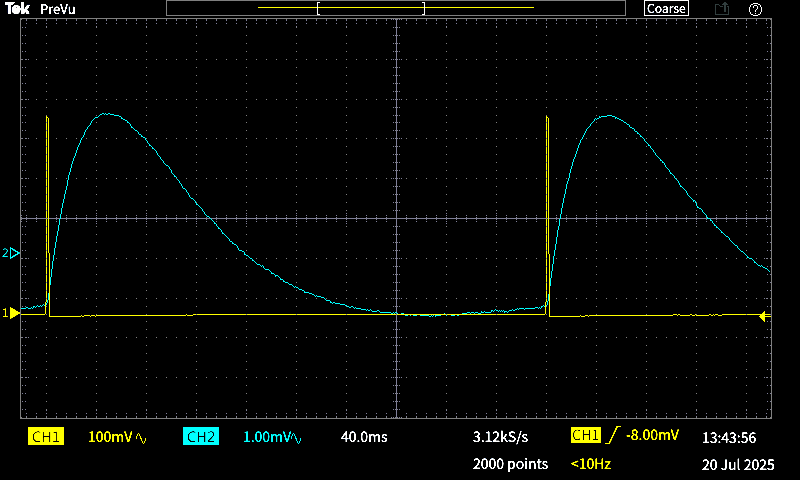}
    \caption{DSO screenshot of the analog filter output for a two-pulse test: two square pulses (each $2$\,ms, approximate bandwidth $\approx 500$\,Hz) separated by $300$\,ms.}
    \label{fig:HW_dso}
\end{figure}

\subsection{Numerical mapping and prototype} To validate the pipeline, we jointly optimized the poles and encoder for $L=2$ with parameter ranges as in \eqref{eq:a_tau_range}, using $T_s = 0.0476$ and $N = 21$. The learned poles were
\[
\alpha_1 = 13.23,\qquad \alpha_2 = 24.44.
\]
Fixing convenient capacitor values $C_1=C_2=1\,\mu\mathrm{F}$, the ideal resistor sum and product implied by \eqref{eq:ckt_param} are
\[
R_1 + R_2 = \frac{1/\alpha_1 + 1/\alpha_2}{C_2},\qquad R_1 R_2 = \frac{1}{\alpha_1\alpha_2 C_1 C_2}.
\]
Solving the resulting quadratic yields ideal resistor values (numerical details in the companion appendix). For the prototype we selected nearby practical values, $R_1 = 85\ \mathrm{k}\Omega$ and $R_2 = 36.5\ \mathrm{k}\Omega$, with $C_1=C_2=1\,\mu\mathrm{F}$. The implemented poles shift modestly from the learned ones (relative deviations on the order of $\sim$10\%), but the encoder still recovers locations accurately.

\subsection{Bench validation and results} 
\begin{figure}[!t]
        \centering
        \includegraphics[width=\linewidth]{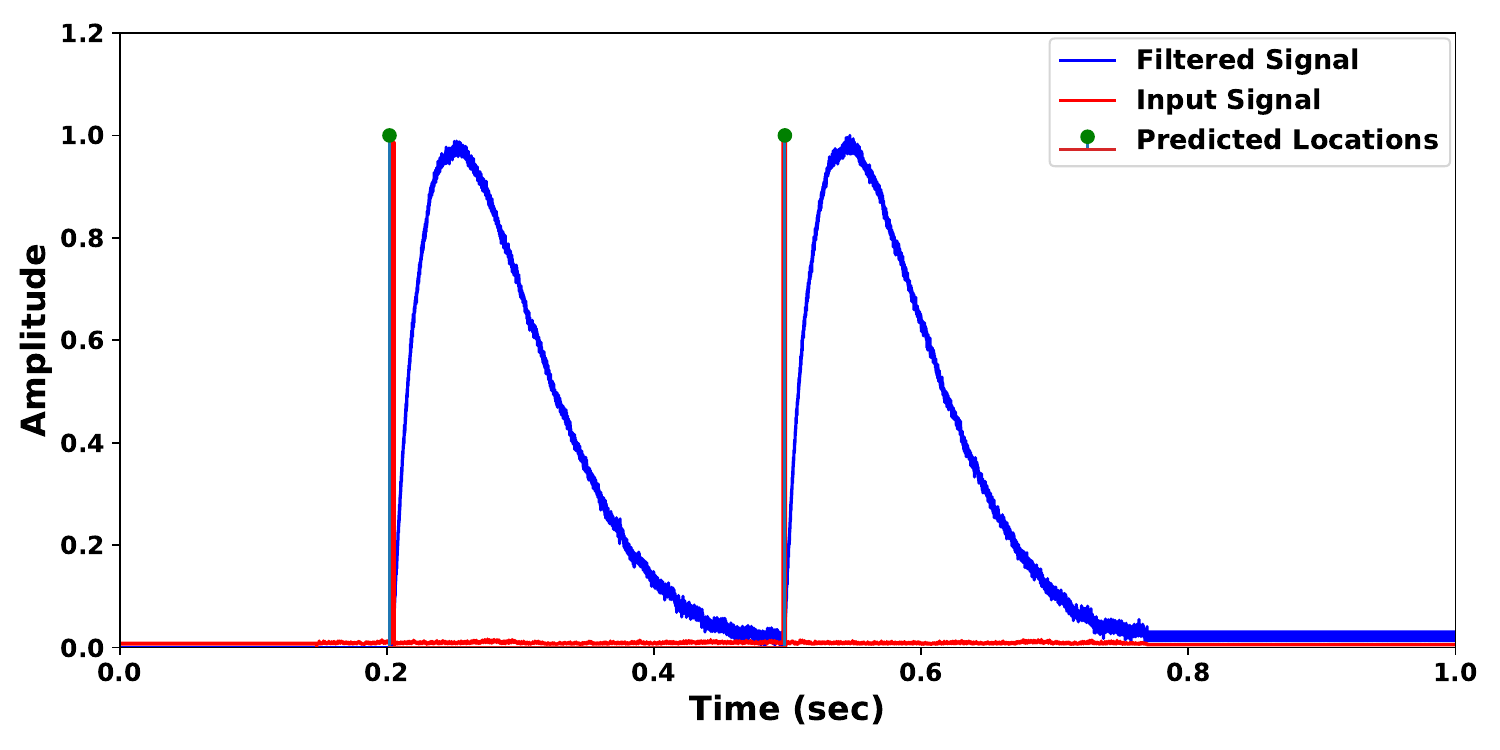}
        \caption{Hardware signal with predicted and true locations.}
        \label{fig:output}
\end{figure}

The implemented filter was tested on the bench setup shown in Fig.~\ref{fig:HW_stp}. Input signals were $L=2$ square pulses (duration $2\,$ms, separation $300\,$ms); the filtered output was captured on a DSO (Fig.~\ref{fig:HW_dso}) and sampled at $200\,$Hz (one-fifth of the Nyquist rate for the pulse bandwidth). The encoder processed these samples to estimate pulse locations and amplitudes. A representative trial (shown in Fig.~\ref{fig:output}) produced true pulse locations $[0.2,0.5]$ and estimates $[0.202,0.498]$, that is, absolute timing errors of $2$\,ms per pulse and an NMSE of approximately $-45\,$dB. These results closely match the simulations and validate that the learned two-exponential kernels are implementable in analog hardware.

A few quick remarks related to the practical implementation are in order.  
\begin{itemize}
  \item \textbf{Rounding and selection:} We selected standard-valued components nearest to the ideal solutions to avoid impractically large/small values; this introduces pole shifts (by $10\%$) but did not materially degrade reconstruction.
  \item \textbf{Optimization constraints:} Simple bounds on $\alpha_i$ were enforced to avoid extreme pole values that would map to impractical RC ranges or demand very high-Q analog stages.
  \item \textbf{Sources of discrepancy:} Remaining mismatch between simulation and bench results is primarily due to ADC quantization, component tolerances, op-amp finite bandwidth/slew-rate, and measurement-chain imperfections.
\end{itemize}


In summary, parametrizing the sampling kernel as a sum of exponentials permits a direct mapping from learned poles to realizable RC filter networks, and the prototype demonstrates that this mapping yields practically useful recovery with low-rate sampling.

\section{Conclusion}
\label{sec:conclusion}
In this paper, we show that accurate FRI reconstruction does not require sum-of-exponential–generating kernels: by jointly learning a sampling kernel and a CNN encoder, our pipeline delivers resolution-aware, noise-robust recovery while operating at substantially lower sampling rates than FRIED-Net under matched conditions. Starting from known truncated-Gaussian and Gaussian-pair kernels, we demonstrate stable performance; with joint learning, the kernel adapts to data and yields consistent NMSE gains across SNRs and resolutions, remaining competitive even with reduced samples (\(N=11\)) and higher model orders (\(L=5,10\)). The learned kernel is practically realizable as a second-order, unity-gain Sallen--Key filter via optimization of two exponential poles, and hardware measurements closely track simulations, underscoring deployability. Overall, this work advances the FRI paradigm by unifying kernel design and estimation into a single data-driven framework that is resolution-aware, noise-tolerant, and directly implementable in hardware.

\bibliographystyle{IEEEtran}
\bibliography{refs,refs2,SoE_learn}

\end{document}